\documentclass[a4paper,10pt]{article}

\usepackage{amssymb,amsfonts}

\newcommand{\be}{\begin{eqnarray}}
\newcommand{\ee}{\end{eqnarray}}
\newcommand{\Tr}{\,{\rm Tr}\,}

\newtheorem{theorem}{Theorem}[section]

\def\Mo{{\mathbb M}}
\def\Co{{\mathbb C}}
\def\Do{{\mathbb D}}
\def\Io{{\mathbb I}}
\def\Xo{{\mathbb X}}

\begin{document}

\title{
Extension of Information Geometry to Non-statistical Systems: Some Examples
}

\author{Jan Naudts and Ben Anthonis\\
Universiteit Antwerpen\\
}

\maketitle

\begin{abstract}
Our goal is to extend information geometry to situations where statistical modeling  is not obvious.
The setting is that of modeling  experimental data.
Quite often the data are not of a statistical nature.
Sometimes also the model is not a statistical manifold.
An example of the former is the description of the Bose gas in the grand canonical ensemble.
An example of the latter is the modeling  of quantum systems with density matrices.
Conditional expectations in the quantum context are reviewed.
The border problem is discussed: through conditioning the model point shifts to the border of the differentiable manifold.

\end{abstract}

\section{Introduction}

One of the goals of information geometry \cite{AN00} is the study of the geometry of a statistical manifold $\Mo$.
A tool suited for this study is the divergence function $D(p||q)$, called relative entropy in the physics literature.
It compares two probability distributions $p$ and $q$. It cannot be negative and vanishes if and only if $p=q$.
In our recent works \cite{NA12,NA13,AB14} we have stressed the importance of
considering the statistical manifold $\Mo$ as embedded in the set of all probability distributions.
In particular, the divergence function $D(p||q)$, with $p$ not belonging to $\Mo$, can be used to characterize
exponential families. We also stressed that it is not a strict necessity that the first argument
of the divergence is a probability distribution. The first example given in this paper is an illustration
of this point.

In quantum probability \cite{AL81,PD08} both arguments of the divergence function are replaced by density matrices,
which are the quantum analogues of probability distributions. A renewed interest in quantum probability
comes from quantum information theory (see for instance \cite{PD08,NC00}). The theoretical developments are
accompanied by a large number of novel experiments. Some of them are mentioned below \cite{AGR82,PGMGS01,GBDSGKBRS07}.
They confirm the validity of quantum mechanics but challenge our understanding of nature.
The present paper tries to situate some recent insights in the context of quantum conditional expectations.
In particular, weak measurement theory \cite{AAV88,AV07} is considered.

\section{The Ideal Bose Gas}

The result of a thought experiment on the ideal Bose gas is a sequence of non-negative integers
$n_1,n_2,\cdots$ with finite sum $\sum_{j=1}^\infty n_j<+\infty$.
A model for these data is a two-parameter family of probability distributions
\be
p_{\beta,\mu}(n)&=&\frac{1}{Z(\beta,\mu)}\exp(-\beta\sum_j\epsilon_jn_j+\beta\mu\sum_jn_j),
\label {ibg:pdf}
\ee
with normalization given by
\be
Z(\beta,\mu)=\prod_j\frac{1}{1-\exp(-\beta(\epsilon_j-\mu))}.
\ee
The numbers $\epsilon_j$ are supposed to be known and to increase fast enough with the index
$j$ to guarantee the convergence of the infinite product. The parameters $\beta$ and $\mu$
are real. By assumption is $\beta>0$ and $\mu<\epsilon_j$ for all $j$.

The model space is the statistical manifold $\Mo$ 
formed by the probability distributions $p_{\beta,\mu}$.
However, the data produced by the measurement
are strictly spoken not of a stochastic origin. 
Indeed, a more detailed modeling  of the experiment involves
quantum mechanics and quantum measurement theory. The latter is much debated since the
introduction of so-called weak measurements \cite{AAV88}.
Hence, we cannot speculate about a possible stochastic origin of the data.

If all we know about the experiment is that it produces the sequence $n$ then
the simplest modeling  we can do is to fit (\ref {ibg:pdf}) to the data
with a method which can be used also outside the conventional settings of statistics.
Our proposal is to do the fitting by minimization of a divergence function.

The Kullback-Leibler divergence between two probability measures can be easily generalized
to a divergence between a sequence of integers $n$ and a point $(\beta,\mu)$ of the statistical manifold $\Mo$.
Our ansatz is
\be
D(n||\beta,\mu)&=&\ln Z(\beta,\mu)-\sum_jn_j(-\beta\epsilon_j+\beta\mu).
\label{ibg:div}
\ee
Minimizing this divergence produces a best fit for the data.
If such a best fit $\beta,\mu$ exists we say that it is the orthogonal projection of $n$
on $\Mo$. The reference to orthogonality is justified by the knowledge that a
Pythagorean theorem holds for the divergence (\ref{ibg:div}) --- see \cite {AB14}.
Let us analyze in what follows the properties of this minimization procedure.

Derivatives of (\ref {ibg:div}) w.r.t.~$\beta$ and $\mu$ can be calculated easily.
It follows that, if the minimization procedure has a solution, then
it satisfies the pair of equations
\be
\sum_jn_j\epsilon_j&=&\sum_j\frac{\epsilon_j}{\exp(\beta(\epsilon_j-\mu))-1},\\
\sum_jn_j&=&\sum_j\frac{1}{\exp(\beta(\epsilon_j-\mu))-1}.
\ee
These expressions are well-known in statistical physics, see for instance \cite {HK01}, Chap.~10.

A metric tensor $g(\beta,\mu)$ is given by the matrix of second derivatives of\\
$D(n||\beta,\mu)$, evaluated at the minimum. One finds
\be
g(\beta,\mu)&=&\sum_j\frac{\exp(\beta((\epsilon_j-\mu))}{[\exp(\beta(\epsilon_j-\mu))-1]^2}
\left(
\begin{array}{lr}
\epsilon_j-\mu)^2 &-\beta(\epsilon_j-\mu)\\
-\beta(\epsilon_j-\mu) &\beta^2
\end{array}
\right).
\ee
It is positive definite and does not depend on the choice of coordinates $\beta,\mu$
of the statistical manifold $\Mo$, as it should be.

The next step is the introduction of covariant derivatives $\nabla_a$, $a=\beta,\mu$
such that  for all $n$ the Hessian of the divergence $D(n||\beta,\mu)$,
evaluated at the projection point  $(\beta,\mu)$ of $n$ on $\Mo$, equals the Hessian of a potential
$\Phi(\beta,\mu)$. 
The corresponding connection $\omega$ satisfies
\be
\nabla_a\partial_b=\omega^c_{\,ab}\partial_c.
\ee
The existence of this connection $\omega$ shows that the Hessian $\nabla_a\nabla_b D(n||\beta,\mu)$
is constant on the set of all $n$ which project on the point $(\beta,\mu)$
and equals the metric tensor $g$. This gives the inverse of $g$ the meaning of a Fisher information.

A method for calculating $\omega$ is given in \cite{AB14}.
It turns out that all coefficients of $\omega$ vanish except $\omega^\mu_{\,\beta\mu}=1/\beta$.

\section{Quantum Measurements}

The quantum analogue of a probability distribution is a density matrix. In the finite-dimensional case
this is a positive-definite matrix whose trace equals 1.
Its eigenvalues $\lambda_j$ satisfy $\lambda_j\ge 0$ and $\sum_j\lambda_j=1$.
Hence they can be interpreted as probabilities.

On the other hand the state of the quantum system, in the most simple case, is described by
a wave function $\psi$. This is a normalized element of a Hilbert space
${\cal H}$. Let $|\psi\rangle\langle\psi|$ denote the orthogonal projection onto the subspace $\Co\psi$.
This is a density matrix of rank 1. It is generally believed that a measurement on the
quantum system with wave function $\psi$ yields the diagonal part of the matrix 
$|\psi\rangle\langle\psi|$ in an orthonormal basis the choice of which is dictated by the experimental setup.
Let $(e_j)_j$ denote this basis. Then the measured quantities are the numbers
$|\langle e_j|\psi\rangle|^2$ (here $\langle\cdot|\cdot\rangle$ is the inner product of the Hilbert space).
These are the diagonal elements of the matrix $|\psi\rangle\langle\psi|$.
The diagonal part of this projection operator is again a density matrix, which we denote
$\mbox{diag}(|\psi\rangle\langle\psi|)$.
It is the result of the experiment.

The map
\be
E:\,|\psi\rangle\langle\psi|\rightarrow \mbox{diag}(|\psi\rangle\langle\psi|)
\ee
can be seen as a conditioning which is introduced by the experimental setup.
Indeed, $E$ is a quantum conditional expectation in the terminology of Petz (Chap.~9 of \cite {PD08}).
See the Appendix \ref {sect:appA}).
Petz gives an overview of quantum probability theory as it is known today.
The part on conditional expectations originated with the work of 
Accardi and Cecchini \cite{AC82} and relies on Tomita-Takesaki theory.

We are interested in the question how the conditioning interferes with the 
modeling  of experimental data using a divergence function.
The quantum analogue of the Kullback-Leibler divergence (also called the relative entropy)
has density matrices as its arguments. It is given by
\be
D(\sigma||\rho)&=&\Tr\sigma\ln\sigma-\Tr\sigma\ln\rho.
\label{quant:KL}
\ee
Let $\Xo$ denote the set of density matrices $\sigma$ for which $\sigma\ln\sigma$
is trace-class. Let $V_\sigma$ denote the set of density matrices $\rho$
such that ${\cal R}(\sigma)\subset {\cal R}(\rho)$ and $\sigma\ln\rho$ is a trace-class
operator. The domain of $D$ is then
\be
\Do=\{(\sigma,\rho):\,\sigma\in\Xo,\rho\in V_\sigma\}.
\ee

For the sake of completeness we repeat here the following well-known result
(see Theorem 5.5 of \cite{OP93})

\begin{theorem}
\label{thm1}
 $D(\sigma||\rho)\ge 0$, with equality if and only if $\sigma=\rho$.
\end{theorem}
\noindent

Fix now a model manifold $\Mo$.
It is tradition to work with the
quantum analogue of a Boltzmann-Gibbs distribution, which is a probability
distribution belonging to the exponential family (see for instance \cite{NJ11}).
The parametrized density matrix $\rho_\theta\in\Mo$ is of the form
\be
\rho_\theta&=&\frac{1}{Z(\theta)}e^{-\theta^kH_k},
\ee
with normalization
\be
Z(\theta)&=&\Tr e^{-\theta^kH_k}.
\ee
The operators $H_k$ are self-adjoint. Together they form the Hamiltonian of the system
under consideration. The parameters $\theta^1,\theta^2,\cdots,\theta^n$ are real numbers.
Note that Einstein's summation convention is used.

The estimation problem is the question about the optimal choice of the parameters $\theta^k$
given the result $\sigma$ of the experiment.
The proposal of information geometry is to use the divergence function (\ref {quant:KL})
to calculate the orthogonal projection $\rho_\sigma$ of $\sigma$ onto the model manifold
$\Mo=\{\rho_\theta:\,\theta\in\Theta\}$. The projection is said to be orthogonal because the following
Pythagorean relation holds
\be
D(\sigma||\rho_\theta)&=&D(\sigma||\rho_\sigma)
+D(\rho_\sigma||\rho_{\theta})
\ee
holds for all $\theta$ in $\Theta$.

Assume now that an experiment is done in the basis $(\psi_n)_n$ in which the
elements of $\Mo$ are diagonal.
Let $\sigma_c\equiv\mbox{diag}(\sigma)$ as before.
Then it follows from Theorem 9.3 of \cite{PD08} that
\be
D(\sigma||\rho)&=&D(\sigma||\sigma_c)+D(\sigma_c||\rho)
\quad\mbox{ for all }\rho\in \Mo.
\label{qm:petzrel1}
\ee
Now fitting the result $\sigma_c$ of the experiment with elements of $\Mo$
is equivalent with fitting the unknown density matrix $\sigma$
because the difference of the two divergences is constant, equal to $D(\sigma||\sigma_c)$.

\section{Weak Measurements}

In many recent experiments the actual state of the system, which is described by the density matrix $\sigma$,
is measured in a basis $(\psi_n)_n$ in which $\sigma$ is far from diagonal.
Many of these experiments involve so-called quantum entangled particles.
They confirm \cite {AGR82} that the Bell inequalities,
which are derived using probabilistic arguments (see for instance \cite {CHSH69}),
can be violated. 

Because one knows that the actual state $\sigma$ is not diagonal one tries to fit a model
which is not diagonal as well. In such a case the above argument based on (\ref {qm:petzrel1})
cannot be used. Instead, the conditioning implied by the experimental setup should be included
in the modeling of the experiment.

Introduce a conditional manifold
\be
\Mo_c=\{\rho_c:\,\rho\in\Mo\mbox{ and }\rho>0\},
\quad\mbox{ where }\rho_c\equiv \mbox{diag}(\rho).
\label{qm:condman}
\ee
The relation (\ref {qm:petzrel1}) then shows that the optimal $\rho_c$,
minimizing the divergence $D(\sigma_c||\rho_c)$, also minimizes  $D(\sigma||\rho_c)$.

It can happen\footnote {If $\Mo_c$ is empty there is not much to tell.}
that $\Mo_c$ is in the border region of the manifold of positive-definite matrices,
where the value of the function $\rho_c\rightarrow D(\sigma||\rho_c)$ can become very large.
This is similar to the effect exploited in weak measurements \cite{AAV88},
namely that the denominator of the so-called weak value can become very small.
See the Appendix \ref {sect:app:wm}. This suggests
that weak measurements can be understood by the behavior of the divergence
function $\rho_c\rightarrow D(\sigma||\rho_c)$ in the border region.
This idea requires further exploration.

In the more common von Neumann type of experiments the measurement disturbs
the quantum system in such a strong manner that the conditioning of the experimental
setup also changes the state of the quantum system. Repeating the experiment
then reproduces the same outcome as that of the first measurement.
This is called the collapse of the wave function.
If the outcome $\rho_c$ of the experiment is very sensitive to small changes
in the state $\sigma$ of the quantum system then one can afford to make the
interaction between quantum system and measurement apparatus so weak that
repeated measurements become feasible. In recent experiments \cite {GBDSGKBRS07} 
thousands of consecutive measurements were feasible.
They reveal a gradual change of the quantum state $\sigma$
of the system as a consequence of the measurements.

\section{Summary}

Two situations are described where a divergence function is used with
arguments which are {\sl not} probability distributions.
In the example of the ideal Bose gas the experimental data are sequences $n=(n_j)_j$
of non-negative integers. It is more natural to take the sequence $n$ as the first
argument of the divergence function rather than to introduce an empirical measure
concentrating on the data points. In the example of quantum mechanics the arguments
are density matrices. The use of density matrices as the arguments of the divergence
has been  studied extensively in the context of quantum probability.

In the final part of the paper we investigate the use of divergences in the theory
of quantum measurements. Our point of view is that any quantum measurement
necessarily induces a quantum condition on the range of experimental outcomes.
The mathematical notion of a quantum conditional expectation is used --- see the Appendix  \ref {sect:appA}.
The recent development of weak quantum measurements is cast into this terminology.
The distinction is made between the conditioning of the experimental outcomes,
which is unavoidable, and the conditioning of the actual state of the system, which
is avoided by the weak measurements. The eventual importance of the border
of the manifold of positive definite density matrices is pointed out.

\appendix

\section{Quantum Conditional Expectations}
\label{sect:appA}

Following Petz (see Chapter 9 of \cite {PD08})
a {\bf conditional expectation} consists of 
a subalgebra ${\cal A}$ of the algebra ${\cal B}$ of bounded linear operators in the Hilbert space $\cal H$
together with a linear map $E:\,{\cal B}\rightarrow {\cal A}$.
They should satisfy
\begin{itemize}
\item $\Io$ belongs to $\cal A$ and $E(\Io)=\Io$.
\item If $A\in{\cal A}$ then also $A^\dagger\in{\cal A}$.
\item If $B$ is positive then also $E(B)$ is positive.
\item $E(AB)=AE(B)$ for all $A\in{\cal A}$ and $B\in{\cal B}$.
\end{itemize}
Take $B=\Io$ in the latter to find that
$E(A)=A$ for all $A$ in ${\cal A}$.

In the terminology of \cite {PD08} a density matrix $\rho$ is {\bf preserved} by the
conditional expectation ${\cal A},E$ if 
\be
\Tr\rho B=\Tr\rho E(B)
\ee
holds for all $b$ in $\cal B$.

Now, let be given an orthonormal basis $(\psi_n)_n$ of $\cal H$.
Then any bounded operator $B$ has matrix elements
$(\langle\psi_m|B\psi_n\rangle)_{m,n}$.
The diagonal part of the operator $B$ is then defined by linear extension of
\be
\mbox{diag}(B)\psi_n&=&\langle\psi_n|B\psi_n\rangle \psi_n.
\ee
The map $B\rightarrow\mbox{diag}(B)$, together with the algebra of all diagonal operators
is a conditional expectation. In addition, for any density matrix $\rho$ the diagonal part
$\rho_c\equiv\mbox{diag}(\rho)$ is again a density matrix and it is preserved by this conditional expectation.
Indeed, one has for all $\rho$ and $B$
\be
\Tr\,\rho_cB&=&\Tr\,\rho_c\mbox{diag}(B).
\ee

\section{Weak Measurement Theory}
\label{sect:app:wm}

In the seminal paper \cite{AAV88} about weak measurements an experimental setup is proposed.
The quantum system contains two parts. The first part is the system  of interest.
It is weakly coupled to the second part. On the latter von Neumann type measurements
are performed to collect data. The subsequent experimental implementations follow
the same scheme. See for instance \cite {PGMGS01,GBDSGKBRS07}.
In the present paper only the first part of the experimental setup is considered as the quantum system.
The remainder is then considered to be part of the measuring apparatus.

Ref.~\cite {AV07} discusses the notions of pre and post selected states.
The density matrix $\sigma=|\psi\rangle\,\langle\psi|$ of the present paper describes
the preselected state. It is the initial state of the experiment transported forward in time
using the Schr\"odinger equation.
In a von Neumann type of measurement the post selected state is the state $|\psi_f\rangle\,\langle\psi_f|$
obtained after the collapse of the wave function, transported backwards in time to the point where it meets
the preselected state.
The claim of \cite {AV07} is that the result of the measurement is a so-called
weak value of an operator $C$, which is the operator of the quantum system 
to which the measurement apparatus couples.
This weak value is given by
\be
\langle C\rangle&=&\frac{\langle\psi_f|C\psi\rangle}{\langle\psi_f|\psi\rangle}.
\ee
It can become arbitrary large by setting up the experiment in such a way that 
the overlap $|\langle\psi_f|\psi\rangle|^2$ of the pre and post selected states is very small.
This theory of weak measurements 
has been criticized in the literature (see the references in  \cite {AV07}).
Additional experiments are needed for its validation.

In the terminology of the present paper the coupling via the operator $C$ induces a conditioning
on the outcomes of the experiment.

\end{document}